\documentclass{llncs}
\usepackage{graphicx}
\usepackage[table]{xcolor}
\usepackage{listings} 
\usepackage{cite}

\usepackage{booktabs}
\usepackage{multirow}

\usepackage{hyperref} 

\begin{document}
\title{A Unified Nanopublication Model for Effective and User-Friendly Access to the Elements of Scientific Publishing}

%
\author{Cristina-Iulia Bucur\inst{1}\orcidID{0000-0002-7114-6459} \and
Tobias Kuhn\inst{1}\orcidID{0000-0002-1267-0234} \and
Davide Ceolin\inst{3}\orcidID{0000-0002-3357-9130}}
\authorrunning{C.I. Bucur et al.}
%
\institute{Vrije Universiteit Amsterdam, Amsterdam, The Netherlands 
\email{\{c.i.bucur,t.kuhn\}@vu.nl} \and
Centrum Wiskunde \& Informatica, Amsterdam, The Netherlands
\email{davide.ceolin@cwi.nl}}
\maketitle              
\begin{abstract}
Scientific publishing is the means by which we communicate and share scientific knowledge, but this process currently often lacks transparency and machine-interpretable representations.
Scientific articles are published in long coarse-grained text with complicated structures, and they are optimized for human readers and not for automated means of organization and access. Peer reviewing is the main method of quality assessment, but these peer reviews are nowadays rarely published and their own complicated structure and linking to the respective articles is not accessible.
In order to address these problems and to better align scientific publishing with the principles of the Web and Linked Data, we propose here an approach to use nanopublications as a unifying model to represent in a semantic way the elements of publications, their assessments, as well as the involved processes, actors, and provenance in general.
To evaluate our approach, we present a dataset of 627 nanopublications representing an interlinked network of the elements of articles (such as individual paragraphs) and their reviews (such as individual review comments). Focusing on the specific scenario of editors performing a meta-review, we introduce seven competency questions and show how they can be executed as SPARQL queries.
We then present a prototype of a user interface for that scenario that shows different views on the set of review comments provided for a given manuscript, and we show in a user study that editors find the interface useful to answer their competency questions.
In summary, we demonstrate that a unified and semantic publication model based on nanopublications can make scientific communication more effective and user-friendly.
\end{abstract}
\section{Introduction} \label{introduction}
Scientific publishing is about how we disseminate, share and assess research. Despite the fact that technology has changed how we perform and disseminate research, there is much more potential for scientific publishing to become a more transparent and more efficient process, and to improve on the age-old paradigms of journals, articles, and peer reviews \cite{Priem2013, Berners-Lee2001}. 
With scientific publishing often stuck to formats optimized for print such as PDF, we are not using the advances that are available to us with technologies around the Semantic Web and Linked Data \cite{Clark2014, Sikos2017}.




In this work we aim to address some of these problems by looking at the scientific publishing process at a more finer-grained level and recording formal semantics for the different elements. Instead of treating big bulks of text as such, we propose to represent them as small snippets --- e.g. paragraphs --- that have formal semantics attached and can be treated as independent publication units. They can link to other such units and therefore form a larger entity --- such as a full paper or review --- by forming a complex network of links.
With that approach, we can ensure that provenance of each snippet of information can be accurately tracked together with its creation time and author, and therefore allow for more flexible and more efficient publishing than the current paradigm. A process like peer-reviewing can then be broken down into small snippets and thereby take the specialization of reviewers and the detailed context of their review comments into account, and these review comments can formally and precisely link to exactly the parts of the paper they address.
Each article, paragraph and each review comment thereby forms a single node in a network and is each identified by a dereferenceable URI.

We demonstrate here how we can implement such a system with the existing concept and technology of nanopublications, a Linked Data format for storing small assertions together with their provenance and meta-data. We then show how this approach allows us to build powerful and user-friendly interfaces to aggregate and access larger numbers of such small communication elements, and we demonstrate this on the concrete case of a system for editors to assess manuscripts based on a set of review comments.

In this research we aim to answer the following research questions:
\begin{enumerate}
\item Can we use nanopublications as a unifying data model to represent the structure and links of manuscripts and their assessments in a precise, transparent, and provenance-aware manner?

\item Is a fine-grained semantic publishing and reviewing model able to provide us with answers to common competency questions that journal editors face in their work as meta-reviewers?

\item Can we design an intuitive and effective interface based on a fine-grained semantic publishing and reviewing model that supports journal editors in judging the quality of manuscripts based on the received reviews?
\end{enumerate}
%
We address these research questions with the following contributions:
\begin{itemize}
\item A general scheme of how nanopublications can be used to represent and publish different kinds of interlinked publication elements
\item A dataset of 627 nanopublications, implementing this scheme to represent exemplary articles and their open reviews
\item A set of seven competency questions for the scenario of journal editors meta-reviewing a manuscript, together with SPARQL representations of these questions
\item A prototype of a fine-grained semantic analysis interface for the above scenario and dataset, powered by nanopublications
\item Results from a user study on the perceived importance of the above competency questions and the perceived usefulness of the above prototype for answering them
\end{itemize}

The rest of this article is structured as follows. In Section \ref{background} we describe the current state of the art in the field of scientific publishing and the reviewing process in particular. In Section \ref{approach} we describe our approach with regard to performing the reviewing process in a fine-grained manner based on nanopublications. In Section \ref{evaluation-design} we describe in detail how we performed the evaluation of our approach, while we report and discuss the results of this evaluation in Section \ref{evaluation-results}. Future work and conclusion of the present research are outlined in Section \ref{conclusion}.

\section{Background} \label{background}

Before we move on to describe our approach, we give here the relevant background on scientific publishing, semantic papers, and the specific concept and technology of nanopublications.



Scientific publishing is at the core of scientific research, which has moved in the last decades from print to online publishing \cite{Stern2019}. It is, however, still mostly following the paradigm from the print age, with narrative articles being published in journals and assessed by peer reviewers, only the printed volumes having been replaced by PDF files that are made accessible via search engines \cite{Lippi2018}. Considering the ever increasing number of articles and the increasing complexity of research methods, this old paradigm of publishing seems to have reached its limit, and scientists are struggling to stay up to date in their specific fields \cite{Landhuis2016}.
Slowly but steadily, these old paradigms are shifting with open access publishing, semantically enriched content, data publication, and machine-readable metadata gaining momentum and importance \cite{Shotton2012, Wang2016}.
Opposition is also growing against the use of impact factor \cite{Garfield1999, Garfield2006, Opthof1997} or h-index as metrics for assessment of the participants in this publication process, and it has been shown that these metrics can be tampered with easily \cite{Seglen1997, Dong2005, Saha2003, Alberts2013}.


Advances in Semantic Web technologies like RDF, OWL, and SPARQL have allowed for the semantic enhancement of scholarly journal articles when publishing data and metadata \cite{Shotton2009_1, Shotton2009_2}. As such, semantic publishing was proposed as a way to make scholarly publications discoverable, interactive, open and reusable for both, humans and machines, and to release them as Open Linked Data \cite{Mirri2017, Sateli2016, Jacob2017}. In order to extract formal semantics from already published papers in an automated manner, sophisticated methods such as the compositional and iterative semantic enhancement method (CSIE) \cite{Peroni2017_1}, conceptual frameworks for modelling contexts associated with sentences in research articles \cite{Angrosh2014} and semantic lenses were developed \cite{diIorio2014}. Furthermore, HTML formats like RASH have been proposed to represent scientific papers that include semantic annotations \cite{Peroni2017_2}, and vocabularies like the SPAR (Semantic Publishing and Referencing) suite of ontologies have been introduced to semantically model all aspects relevant to scientific publishing \cite{Peroni2018}.
These approaches mostly work on already published articles, but it has been argued that scientific findings and their contexts should be expressed in semantic representations from the start by the researchers themselves, in what has been named \emph{genuine semantic publishing} \cite{Kuhn2017_1}. 

In our previous work \cite{Bucur2019}, we applied the general principles of the Web and the Semantic Web to promote this kind of genuine semantic publishing \cite{Kuhn2017_1} by applying it to peer reviews. We proposed a semantic model for reviewing at a finer-grained level called Linkflows and argued that Linked Data principles like dereferenceable URIs using open standards like RDF can be used for publishing small snippets of information, such as an individual review comment, instead of big chunks of text, such as an entire review. These small snippets of text can be represented as nodes in a network and can be linked with one another with semantically-annotated connections, thus forming distributed and semantically annotated networks of contributions. The individual review comments are semantically modeled with respect to what part of the paper they target, whether they are about syntax or content, whether they raise a positive or negative point, and whether they are a suggestion or compulsory, and what their impact on the quality of the paper is. We showed on this model that it is indeed beneficial if we capture these semantics at the source (i.e. the peer reviewer in this case).


Nanopublications \cite{Groth2010} are a specific concept and technology based on Linked Data to publish scientific results and their metadata in small publication units. Each nanopublication has an assertion that contains the main content (such as a scientific finding), and comes with provenance about that assertion (e.g. what study was conducted to derive at the assertion; or which documents it was extracted from) and with publication information about the nanopublication as a whole (e.g. by whom and when it was created). All these three parts are represented in RDF and thereby machine-interpretable.

It has been shown how nanopublications can also be used for other kinds of assertions, including meta-statements about other nanopublications \cite{kuhn2013broadening}, and
in order to make nanopublications verifiable and immutable, \emph{trusty URIs} \cite{kuhn2015making} can be used as identifiers, which include cryptographic hash values that are calculated on the nanopublication's content. A decentralized server network has been established based on this, through which anybody can reliably publish and retrieve nanopublications \cite{Kuhn2016}. In order to group nanopublications into larger collections and versions thereof, index nanopublications have been introduced  \cite{Kuhn2017_2}. With these technologies, small interconnected Linked Data snippets can be published in a reliable, decentralized, provenance-aware manner.

\section{Approach} \label{approach}

Our general approach is to investigate the benefits of using nanopublications as a unifying publishing unit to establish a new paradigm of scientific communication that is better aligned with the principles of the Web and Linked Data. We investigate how such an approach could allow us to communicate in a more efficient, more precise, and more user-friendly manner.


\subsection{Semantic Model and Nanopublications} \label{semantic-model-nanopubs}

\begin{figure}[tb]
	\centering
	\includegraphics[width=0.9\textwidth]{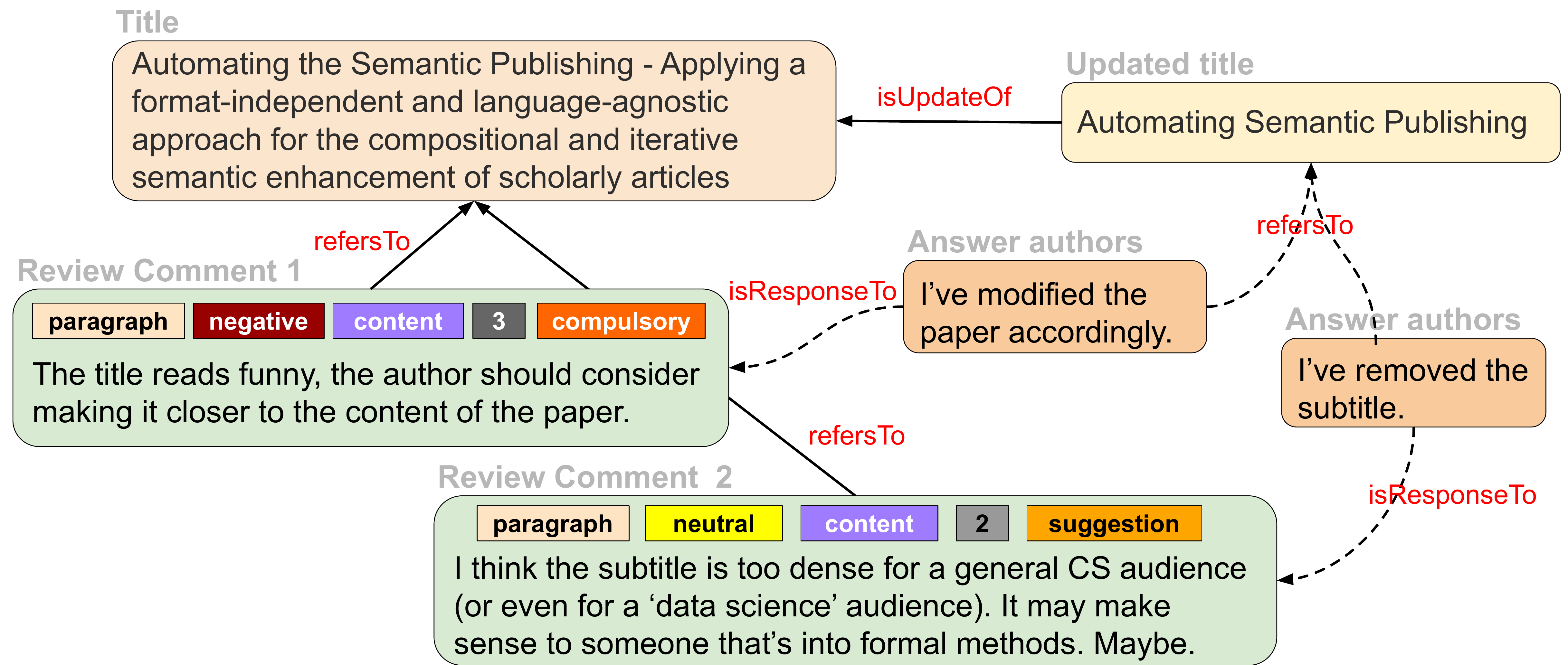}
	\caption{An example of a nanopublication-style communication interaction.}
	\label{fig:diagram-approach}
\end{figure}

Our unifying semantic model based on nanopublications uses a number of existing ontologies like SPAR, PROV-O, FAIR* reviews, the Web Annotation Data Model, and our own Linkflows model \cite{Bucur2019} to break the big bulks of article and review texts into smaller text snippets. An example of a nanopublication-style communication interaction during the reviewing process is illustrated in Figure \ref{fig:diagram-approach}, where the title of a paper is addressed by several review comments that come with semantic classes (e.g. \emph{suggestion}), which are themselves referred to by the authors' answers that link them to the updated version. Each node in this network is represented as a separate nanopublication and all the attributes and relations are formally represented as Linked Data.

\begin{figure}[tb]
\centering
\includegraphics[width=\textwidth]{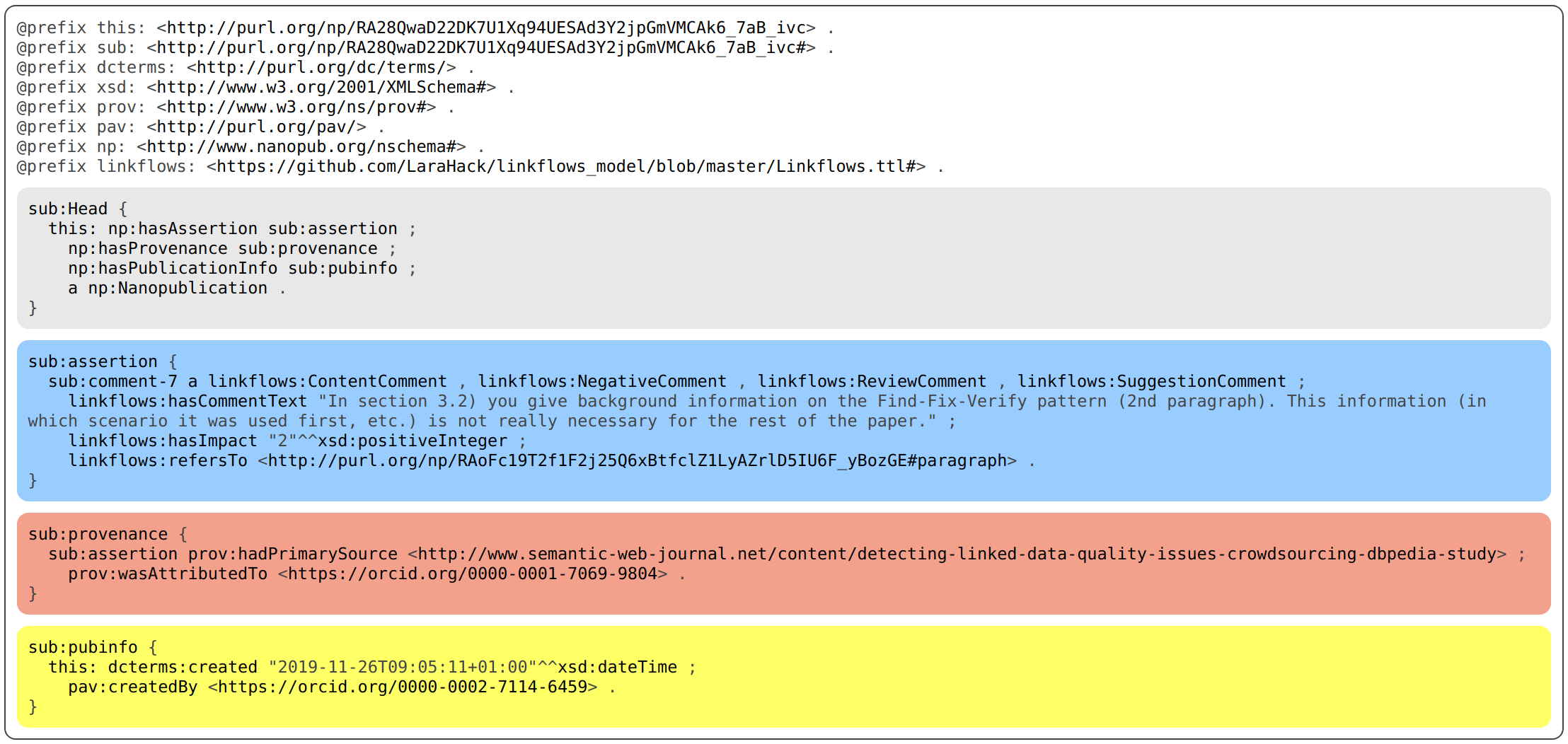}
\caption{Example nanopublication of a review comment.}
\label{fig:example-nanopub}
\end{figure}

As we can see in Figure \ref{fig:diagram-approach}, the properties \textit{refersTo, isResponseTo, isUpdateOf} play the key role of linking different nodes in this network. \textit{refersTo} is a property that links a review comment to the text snippet in the article it refers to. \textit{isResponseTo} links the answer of the authors to the review comments of the reviewer and also to new versions of the text snippets that these review comments triggered.
\textit{isUpdateOf} links a version of the text snippet to another.
%
In our approach, snippets of scientific articles (mostly corresponding to paragraphs) as well as their review comments (corresponding to individual review comments) are semantically represented as nanopublications \cite{Groth2010}, and thereby they each form a node in the network described above. A complete example of such a nanopublication containing a review comment is shown in Figure \ref{fig:example-nanopub}.

Each of the three main parts of a nanopublication --- assertion, provenance, and publication info --- is represented as an RDF graph. In the example of Figure \ref{fig:example-nanopub}, the assertion graph describes a review comment using the classes and properties of the Linkflows model \footnote{\url{https://github.com/LaraHack/linkflows_model}}. It raises a negative point with an importance of 2 out of 5, and is marked as a suggestion for the authors. Furthermore, we see that this review comment refers to an external element, with a URI ending in \path{#paragraph}, as the target of this comment. This external element happens to be a paragraph of an article described in another nanopublication, which we can find out by following that trusty URI link. 

Moreover, the nanopublication contains information regarding the creator of the assertion and the creator of the nanopublication that contains this assertion. These pieces of information can be found in the \textit{provenance} and \textit{publication info} graphs. As illustrated in Figure \ref{fig:example-nanopub}, the author of the review comment is indicated by his ORCID identifier and the source of the original source of the review comment is indicated by the the URL pointing to a link of the Semantic Web Journal. From the publication info graph, we can see who created the whole nanopublication together with the date and time of its creation.


With nanopublications, the provenance and immutability of these small contributions can be guaranteed by the usage of Trusty URIs \cite{Kuhn2014}. As such, for every nanopublication, in order for it to be published, a unique immutable URI is generated to refer to the node that holds the nanopublication. Any change of this nanopublication results in the generation of a new nanopublication, thus of a new node that is linked to the previous one. Such nanopublications can then be published in the existing decentralized nanopublication network \cite{Kuhn2016}.

\subsection{Use Case with Competency Questions} \label{uc-competency-questions}

In the scientific publishing context, editors of journals play a key role, being an important link between content providers for journals (authors), the people who assess the quality of the content (peer reviewers) and the consumers of such content (the readers). While the peer reviewers are the ones that can recommend the acceptance or rejection of an article, it is up to the editors to make the final decision.
We will look here into how our approach can benefit the specific scenario of editors assessing a manuscript based on given reviews and having to write a meta-review.

Performing such a meta-review is not a trivial task. As classical reviews are mainly comprised of a large bulks of text in natural language, it is hard to provide a tool with quantitative information about the reviews and their collective implications on the manuscript. As such, an editor needs to spend a lot of time just to read these reviews fully to even get an overview of the nature and range of the raised issues.

In order to apply our approach to this chosen use case, we first define a set of competency questions (CQs), which are natural language questions that are created with the objective to assess the practicality and coverage of an ontology or model \cite{Bezerra2013}.
After consulting with publishing experts at IOS Press\footnote{\url{https://www.iospress.nl/}} and the Netherlands Institute of Sound and Vision\footnote{\url{https://www.beeldengeluid.nl/en}}, we came up with the following seven quantifiable competency questions from an editor's point of view during meta-reviewing:
\begin{itemize}
\item \textbf{CQ1}: \textit{What is the number of positive comments and the number of negative comments per reviewer?}

\item \textbf{CQ2}: \textit{What is the number of positive comments and the number of negative comments per section of the article?}


\item \textbf{CQ3}: \textit{What is the distribution of the review comments with respect to whether they address the content or the presentation (syntax and style) of the article?}

\item \textbf{CQ4}: \textit{What is the nature of the review comments with respect to whether they refer to a specific paragraph or a larger structure such as a section or the whole article?}

\item \textbf{CQ5}: \textit{What are the critical points that were raised by the reviewers in the sense of negative comments with a high impact on the quality of the paper?}

\item \textbf{CQ6}: \textit{How many points were raised that need to be addressed by the authors, as an estimate for the amount of work needed for a revision?}

\item \textbf{CQ7}: \textit{How do the review comments cover the different sections and paragraphs of the paper?}
\end{itemize}


\subsection{Dataset} \label{dataset}

In order to evaluate our approach on the given use case, we need some data first. For this, we selected three papers that were submitted to a journal that has open reviews (Semantic Web Journal). Therefore, we could also access the full text of the reviews these papers received. We then manually modelled all the article, paragraphs, review comments, their interrelations, as well as their larger structures --- in the form of sections and full articles and reviews --- as individual nanopublications according to our approach. All these elements were thereby semantically modeled, and we could reuse part of our earlier dataset of manually assigned Linkflows categories \cite{Bucur2019}.
Figure \ref{fig:example-nanopub} above shows an example of a nanopublication that resulted from this manual modeling exercise. We would like to stress here that according to the vision underlying our approach, these semantic representations would in the future be generated as such from the start, and therefore this manual effort is only for evaluation purposes.

Apart form nanopublications at the lowest level, such as the one shown in Figure \ref{fig:example-nanopub}, higher-level ones combine them (by simply linking to them) to form larger structures, such as entire sections, papers, and reviews. Section nanopublications, for example, point to their paragraphs and define their order among other metadata. We also created a nanopublication index \cite{Kuhn2017_2} that refers to this set of manually created nanopublications such that we can retrieve and even reuse parts of this dataset for new versions incrementally. All the nanopublications from our dataset are in an online repository\footnote{\url{https://github.com/LaraHack/linkflows_model_implementation}}. 



\subsection{Interface Prototype for Use Case} \label{interface-prototype-uc} 

In order to apply and evaluate our approach on the chosen use case, we developed a prototype of an editor interface that accesses the nanopublications in the dataset presented above to provide a detailed and user-friendly interface to support editors in their meta-reviewing tasks.

\begin{figure}[tb]
	\centering
	\includegraphics[width=0.9\textwidth]{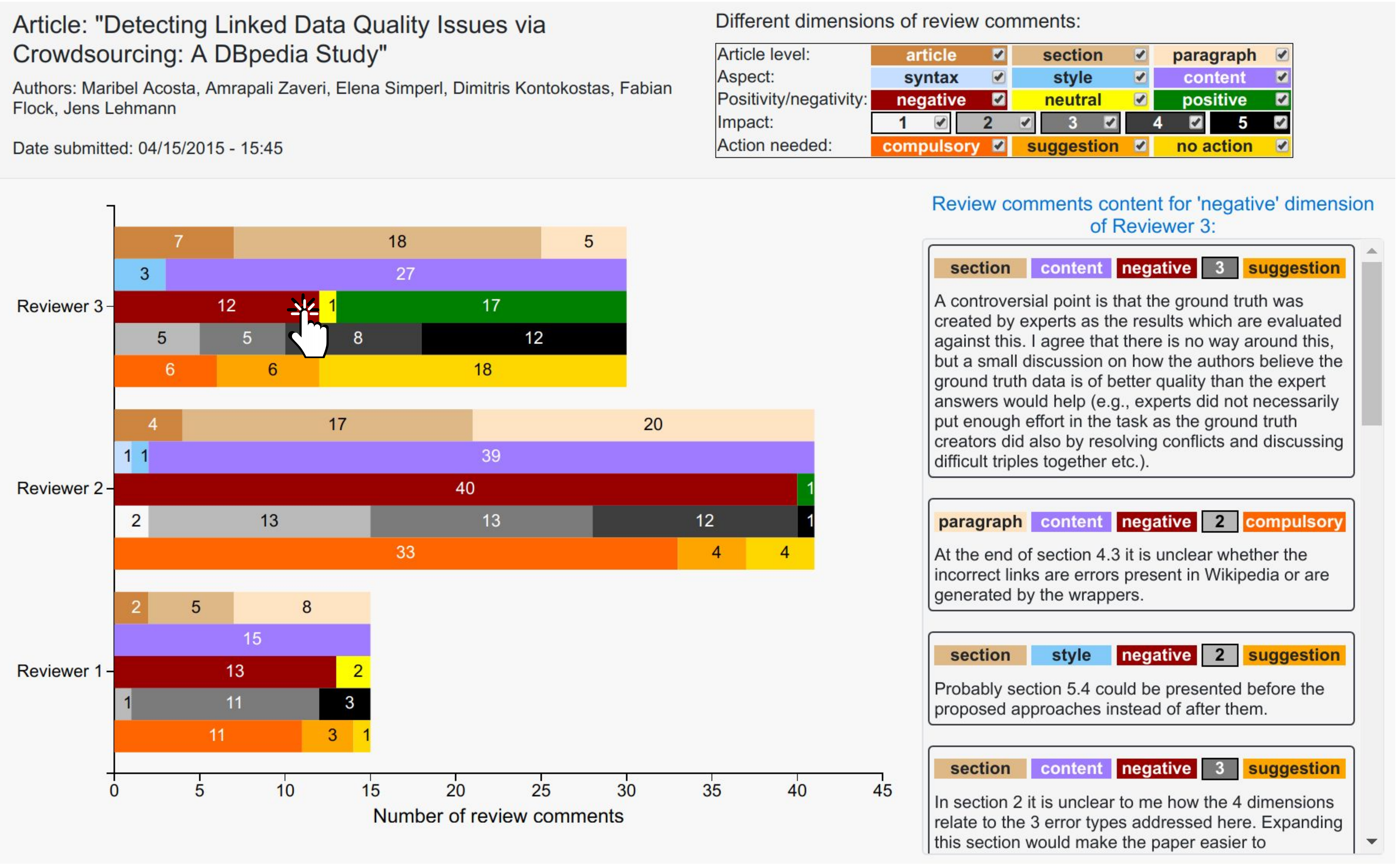}
	\caption{The reviewer-oriented view for the editor study.}
	\label{fig:reviewer-oriented-view}
\end{figure}

\begin{figure}[t]
	\centering
	\includegraphics[width=\textwidth]{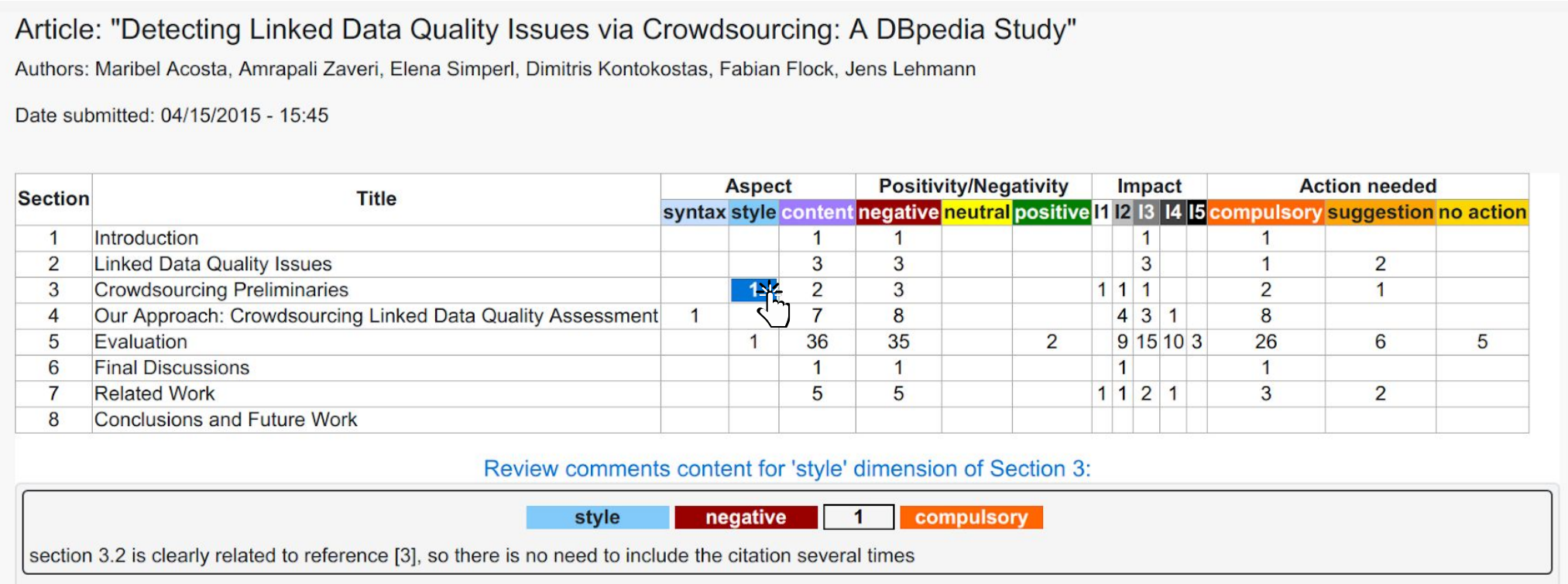}
	\caption{The section-oriented view for the editor study.}
	\label{fig:section-oriented-view}
\end{figure}

This prototype comes with two views: one where the review comments are shown per reviewer in a bar chart broken down into the different dimensions and classes, as shown in Figure \ref{fig:reviewer-oriented-view} and another view that focuses on the distribution of the review comments to the different sections of the article, as shown in Figure \ref{fig:section-oriented-view}. The interface for an exemplary article with three reviews can be accessed online\footnote{\url{http://linkflows.nanopubs.lod.labs.vu.nl}}. The shown content is aggregated from nanopublications stored in a triple store and displayed by showing color codes for the different Linkflows classes for the individual review comments.

In the reviewer-oriented view (Figure \ref{fig:reviewer-oriented-view}), we can see in a more quantitative way the set of review comments and their types represented in different colors, where the checkboxes in the legend can be used to filter the review comments of the given category. To see the content of the review comments that are in a certain dimension, it is sufficient to just click on a bar in the chart.

The section-oriented view (Figure \ref{fig:section-oriented-view}), aggregates all the finer-grained dimensions of the review comments at the level of sections in an article. Again, clicking on one cell in the table, thus selecting one specific dimension of the review comments, will show the content of those review comments underneath the table in the interface.

When data from the triple store is required, the server (implemented in NodeJS with the Express web application framework\footnote{\url{https://nodejs.org}, \url{https://expressjs.com/}}) sends a request
to the Virtuoso triple store where the nanopublications are stored. This request executes a SPARQL query on the stored nanopublications and returns the result to the server that, in turn, passes it further to the client, in the web browser, where the results are postprocessed and visualized. The code for the prototype can be found online\footnote{Interface: \url{ https://github.com/LaraHack/linkflows_interfaces} \\ Backend application: \url{https://github.com/LaraHack/linkflows_model_app} \\ Data: \url{https://github.com/LaraHack/linkflows_model_implementation}}.

\section{Evaluation} \label{evaluation}

Here we present the evaluation of our approach in the form of a descriptive analysis, the analysis of the SPARQL implementations of our competency questions, and a user study with editors on our prototype interface.

\subsection{Evaluation Design} \label{evaluation-design}

First, we run a small descriptive analysis on the nanopublication dataset that we created. We can quantify the size and interrelation of the represented manuscripts and reviews in new ways, including the number of nanopublications, triples, paragraphs, review comments, and links between them. We also tested how long it takes to download all 627 nanopublications from the server network, using \path{nanopub-java} \cite{kuhn2015nanopubjava} as a command-line tool and giving it only the URI of the index nanopublication. This small download test was performed on a personal computer via a normal home network. For this, we retrieved them all via the library's \texttt{get} command and measured the time. We performed this 50 times, in five batches of 10 executions.

Next, we used our dataset to see if we are able to answer the seven competency questions that we defined above, in order to help editors in their meta-reviewing task. With this, we want to find out whether the combination of ontologies and vocabularies we used in our approach is sufficient to cover them, and then whether we can use the SPARQL query language to operationalize them and make them automatically executable on our nanopublication data. 

Finally, we perform a user experiment involving editors to find out whether they indeed consider our competency questions important, and how useful they find our interface for getting an answer to these questions. For this study, we created a form that had two parts corresponding to the two parts of the study. We chose an article from our dataset that had a large number of review comments. For the first part, we asked for the importance of the competency questions using a Likert scale (from 1 to 5). For the second part, we provided static screenshots of our tool (the reviewer-oriented or the section-oriented view, depending on the question) together with a link to the live demo and asked about how useful the participants would find such a tool to answer the given competency question. The answers were on the same kind of a Likert scale from 1 to 5. 
We sent this questionnaire (details online\footnote{\url{https://github.com/LaraHack/linkflows_editor_survey/}}) to a total of 401 editors of journals that support open reviews, specifically Data Science, the Semantic Web Journal and PeerJ Computer Science.

\subsection{Evaluation Results} \label{evaluation-results}

We can now turn to the results of these three parts of our evaluation. Details about the dataset and how it was generated and further queries and results can be found online\footnote{\url{https://github.com/LaraHack/linkflows_model_implementation}}.

\subsubsection{Descriptive Analysis.}  \label{descriptive-analysis}
Our representation of the three papers of our dataset together with their reviews leads to a total of 10\,437 triples in 627 nanopublications, 279 text snippets and 213 review comments (85 for article 1, 59 for article 2 and 69 for article 3). Each of the three articles had three reviews: first article - 17, 18 and 50 review comments provided by the three reviewers, second article - 16, 21, 22 review comments each and third article - 11, 42, 16 review comments each.

In Table \ref{tab:general-statistics-dataset} some general statistics of the dataset are presented, while Table \ref{tab:general-statistics-nanopubs} shows general statistics about the nanopublications corresponding to the three articles and their reviews.
Overall, this demonstrates the working of our approach of representing the elements of scientific communication in a fine-grained semantic manner. Of course, more complex analyses are possible, including network analyses of the complex interaction structure, and the queries for the competency questions that we defined above, to which we come back below.

\begin{table}[tb]
\parbox{.44\linewidth}{
    \centering
    \caption{Descriptive statistics dataset}
    \label{tab:general-statistics-dataset}
    \begin{tabular}{lr}
        \toprule
        part of article & number \\
        \midrule
        articles & 3 \\
        sections & 89 \\
        paragraphs	& 279 \\
        figures	& 11 \\
        tables	& 10 \\
        formula	& 8 \\
        footnote& 2 \\
        review comments	& 213 \\
        \bottomrule
    \end{tabular}
}
\hfill
\parbox{.55\linewidth}{
	\centering
	\caption{Statistics nanopublications.}
	\label{tab:general-statistics-nanopubs}
	\begin{tabular}{lrr}
		\toprule
		 & number & average\\
		\midrule
		Nanopublications: & 627 &  \\
		Head triples: & 2508 & 4.00 \\
		Assertion triples:	& 5420 & 8.64 \\
		Provenance triples:	& 1254 & 2.00\\
		Publication info triples:	& 1255 & 2.00 \\
		\midrule
		Total triples:	&  10\,437 & 16.65 \\
		\bottomrule
	\end{tabular}
	\vspace{5.5mm}
}
\end{table}

Our small test on the performance of retrieving all nanopublications from the decentralized nanopublication network showed an average download time of 11.66 seconds overall (with a minimum of 8.39 and a maximum of 13.34 seconds). This operation retrieves each of the 627 nanopublications separately and then combines them in a single output file. The time per nanopoublication is thereby just 18.6 milliseconds, which is achieved by executing the request in parallel to several servers in the network at the same time.




\subsubsection{Competency Question Execution.}  \label{competency-questions-results}
In order to answer the competency questions in Section \ref{uc-competency-questions}, we managed to implement each of them as a concrete SPARQL query. We can't go into them here in detail due to space limitations, but the complete queries and all the required data and code can be found online\footnote{\url{https://github.com/LaraHack/linkflows_model_implementation/tree/master/queries}}.

This shows that our model is indeed able to capture the needed aspects for our competency questions, but we still need to find out whether these competency questions are indeed considered important by the editors, and whether the results from the SPARQL queries allow us to satisfy these users' information needs. These two aspects are covered in our user study.

\subsubsection{User Study Results.} \label{editor-study-results}
Out of the total 401 questionnaire requests sent, we received a total of 42 answers (10.5\%).
The importance of the seven competency questions for editors and the usefulness of the interface presented to answer these competency questions, assessed on a Likert scale from 1 to 5 where 1 is \textit{not important at all} and 5 is \textit{very important} can be seen in Table \ref{tab:questionnaire-results}. We marked with * the competency questions that had a significant \textit{p}-value ($<$ 0.05) and without, the ones that were not significant. We calculate significance with a simple binomial test by splitting the responses into the ones that assign at least medium importance or usefulness ($\geq3$) and the ones that assign low importance or usefulness ($<3$). 

We see the respondents declared high importance to five of the seven competency questions in a significant manner with average values from 3.05 to 4.58 (CQ1, CQ3, CQ4, CQ5 and CQ6), while the remaining two (CQ2 and CQ7) were not considered important in the editors' view (average values of 2.36 and 2.79, respectively). Apparently, the number of positive and negative comments per section of the article (CQ2) and how the review comments cover the different parts of the article such as sections (CQ7), seem to have mixed reviews from editors, not being considered significantly important. The critical points that were raised by the reviewers (negative comments with a high impact on the paper) seems to be considered the most important competency question for the editors that responded (CQ5) with an average value of 4.58. Also important, in decreasing order, are the distribution of review comments with respect to whether they address the content or the presentation (syntax and style) of the article (CQ3), the number of points raised to be addressed by authors as an estimate for the amount of work needed for a reviewer (CQ6), the number of positive and negative comments per reviewer (CQ1), and the nature of the review comments with respect to whether they refer to a paragraph or a larger structure such as a section or the whole article (CQ4). For CQ2 and CQ7, we can say that editors did find it on average less important which sections of the article the reviews comments addressed. In general, however, we can conclude that most of competency questions are found to be important by most editors. However, we also observe a quite large standard deviation (SD) as seen in Table \ref{tab:questionnaire-results}, ranging from 0.93 to 1.36 on our Likert scale that has a maximum distance of 4.0.

\begin{table}[tb]
    \centering
    \small
    \caption{Results of the user study with editors.}
    \label{tab:questionnaire-results}
    \begin{tabular}{c@{~}|@{~}c@{~}|@{~}c@{~}|@{~}c@{~}|@{~}r@{~}r@{~}|@{~}l@{~}|@{~}r@{~}|@{~}c@{~}|@{~}c@{~}|@{~}r@{~}r@{~}|@{~}l@{~}|}
     & \multicolumn{6}{c@{~}|@{~}}{importance of question} & \multicolumn{6}{c|}{usefulness of interface} \\ \cline{2-13}
     Question & \multirow{2}{*}{AVG} & \multirow{2}{*}{MED} & \multirow{2}{*}{SD} & \multicolumn{2}{c@{~}|@{~}}{count} & $\Delta$count & \multirow{2}{*}{AVG} & \multirow{2}{*}{MED} & \multirow{2}{*}{SD} & \multicolumn{2}{c@{~}|@{~}}{count} & $\Delta$count \\
     &  &  &  & $<$3 & $\geq$3 & $p$-value &  &  &  & $<$3 & $\geq$3 & $p$-value \\
    \hline
    CQ1 & 3.17 & 3 & 1.36 & 15 & 27 & 0.044 * & 3.48 & 4 & 1.17 & 9 & 33 & 1.36e-4 * \\
    CQ2 & 2.36 & 2 & 1.10 & 24 & 18 & 0.860 & 3.83 & 4 & 1.03 & 5 & 37 & 2.22e-7 * \\
    CQ3 & 3.64 & 4 & 0.93 & 5 & 37 & 1.36e-4 * & 3.40 & 3.5 & 1.04 & 9 & 33 & 1.47e-3 * \\
    CQ4 & 3.05 & 3 & 1.19 & 14 & 28 & 0.022 * & 3.26 & 3 & 1.20 & 14 & 28 & 0.022 * \\
    CQ5 & 4.58 & 5 & 0.63 & 0 & 42 & $<$ e-12 * & 3.21 & 3 & 1.16 & 9 & 33 & 1.36e-4 * \\
    CQ6 & 3.57 & 4 & 1.02 & 6 & 36 & 1.41e-6 * & 3.43 & 4 & 1.06 & 8 & 34 & 3.44e-5 * \\
    CQ7 & 2.79 & 3 & 1.12 & 18 & 24 & 0.220 & 3.62 & 4 & 1.03 & 5 & 37 & 2.22e-7 * \\
    \end{tabular}
\end{table}

Next, we evaluated the usefulness of our prototype interface. Here the Likert scale went from 1 standing for \emph{not useful at all} to 5 standing for \emph{very useful}. As we can see from  Table \ref{tab:questionnaire-results}, this interface was on average considered useful for all of the seven competency questions, with averages ranging from 3.21 to 3.83. The preference for scores of 3 or larger is clearly significant for all of them. A substantial minority of respondents, however, didn't find our interface useful leading again to relatively large standard deviation values between 1.06 and 1.19. The free-text feedback field at the end of the questionnaire gave us moreover a variety of suggestions for improvement (some of the editors found the colors too much, others suggested other ways of grouping the data) but without clear overall tendencies.


\section{Discussion and Conclusion} \label{conclusion}

Our results show that we can practically represent the different elements of scientific communication, such as articles and reviews, in a fine-grained and semantic way with nanopublications. We could show that we thereby can automatically answer a wide range of competency questions in the concrete scenario of editors in their meta-reviewing task. We found, however, that some of these were not found to be important, on average, by the editors who participated in our user study. Specifically, the questions about how well the review comments cover the different parts of the paper were not found to be important by a majority of editors. This could indicate that the article structure in terms of its different sections is not a good target for measuring the coverage of reviews. For all the questions, a relatively high variation is observed, which might be hinting at a lack of agreement among editors with respect to how scientific manuscripts should be assessed. This in turn could highlight the importance of more structured and more open reviewing processes. Irrespective of whether the competency questions are important, the majority of editors found our prototype to be useful to answer them, although again with a large variation. With our approach focusing on interoperability and openness, however, it is not necessary to design a single interface that suits everybody, but we could allow editors to choose from several alternatives in the future.

In summary, we could show that nanopublications might be a suitable format not just for scientific findings but also for their reviewing processes. Their open and semantic nature can moreover allow other participants outside of the assigned editor and invited reviewers to contribute with their suggestions and comments, both before and after publication, while all the provenance needed to understand the context of each contribution is recorded. In this way, publication and reviewing as a whole might become more fluid, more inclusive, and more powerful.

\paragraph* \noindent \textbf{Acknowledgements.}
This research was partly funded by IOS Press and the Netherlands Institute for Sound and Vision. The authors would like to thank Stephanie Delbeque, Maarten Fr\"{o}hlich, Erwin Verbruggen, Johan Oomen, and Jacco van Ossenbruggen for providing their insight and expertise.

%
%
%

\bibliographystyle{splncs04}
\bibliography{references.bib}

\end{document}